\documentclass[reprint,amsmath,amssymb, superscriptaddress, aps]{revtex4-2}

\usepackage{float}
\usepackage{comment}
\usepackage{tabularx}

\usepackage[utf8]{inputenc}
\usepackage[T2A,T1]{fontenc}
\usepackage[english]{babel}

\usepackage{color}
\usepackage{ulem}
\usepackage{xcolor}
\usepackage{hyperref}

\definecolor{light_green}{HTML}{bdffc4}
\definecolor{light_blue}{HTML}{bdc9ff}
\definecolor{light_orange}{HTML}{ffebbd}

\usepackage{xfrac}

\usepackage{gensymb}

\usepackage[caption=false]{subfig}

\usepackage{graphicx}
\usepackage{dcolumn}
\usepackage{bm}

\begin{document}

\title{Quadrupole Mie-Resonant Metamaterial}

\author{Ilia M. Fradkin}
\email{I.Fradkin@skoltech.ru}
\affiliation{Skolkovo Institute of Science and Technology, Bolshoy Boulevard 30, bld. 1, Moscow 121205, Russia}
\affiliation{Moscow Institute of Physics and Technology, Institutskiy pereulok 9, Moscow Region 141701, Russia}

\author{Anton V. Nikulin}
\affiliation{Moscow Institute of Physics and Technology, Institutskiy pereulok 9, Moscow Region 141701, Russia}

\author{Nikolay S. Solodovchenko}
\affiliation{Department of Physics and Engineering, ITMO University, St. Petersburg 197101, Russia}

\author{Dmitry S. Filonov}
\affiliation{Moscow Institute of Physics and Technology, Institutskiy pereulok 9, Moscow Region 141701, Russia}
\author{Denis G. Baranov}
\affiliation{Moscow Institute of Physics and Technology, Institutskiy pereulok 9, Moscow Region 141701, Russia}

\author{Mikhail V. Rybin}
\affiliation{Department of Physics and Engineering, ITMO University, St. Petersburg 197101, Russia}
\affiliation{Ioffe Institute, St. Petersburg 194021, Russia}

\author{Kirill B. Samusev}
\affiliation{Department of Physics and Engineering, ITMO University, St. Petersburg 197101, Russia}
\affiliation{Ioffe Institute, St. Petersburg 194021, Russia}

\author{Mikhail F. Limonov}
\affiliation{Department of Physics and Engineering, ITMO University, St. Petersburg 197101, Russia}
\affiliation{Ioffe Institute, St. Petersburg 194021, Russia}

\author{Sergey A. Dyakov}
\affiliation{Skolkovo Institute of Science and Technology, Bolshoy Boulevard 30, bld. 1, Moscow 121205, Russia}

\author{Nikolay A. Gippius}
\affiliation{Skolkovo Institute of Science and Technology, Bolshoy Boulevard 30, bld. 1, Moscow 121205, Russia}

\date{\today}

\begin{abstract}

Dense lattices of photonic crystals can serve as artificial materials, with light propagation in these structures described by effective material parameters that surpass the capabilities of natural materials. In this study, we introduce a metamaterial that supports magnetic quadrupole polarization, a characteristic rarely observed in existing structures. We experimentally demonstrate a magnetic quadrupole metamaterial associated with Mie-resonance-excited stop bands below the Bragg band. Additionally, we develop a theoretical model that addresses both dispersion and boundary conditions within this framework. Using the Fabry-Perot resonator as a case study, we validate our model and reveal that the quadrupole metamaterial can exhibit a markedly different reflection/transmission spectrum, including zero reflection at normal incidence. Our findings underscore the practical potential for both experimental and theoretical investigations of metamaterials that extend beyond the dipole approximation.

\end{abstract}
\maketitle

\section{Introduction}

Rapidly increasing functionality of modern optical devices naturally requires advanced optical materials. In search for new optical properties researchers actively study 2D and van der Waals (especially transition metal dichalcogenide) materials~\cite{Basov2016, Low2017, xia2014two,ermolaev2021giant,ermolaev2024wandering,munkhbat2023nanostructured}), quantum materials~\cite{basov2017towards,keimer2017physics,tokura2017emergent} and new glasses~\cite{alekseev2021multicomponent,alekseev2022local,xu2020ultrahigh,mao2018optical}. Nevertheless, the concept of artificial optical materials or metamaterials~\cite{yablonovitch1987inhibited,engheta2006metamaterials,liu2011metamaterials,simovski2018composite,cui2024roadmap,simovski2012wire,shaposhnikov2023emergent} is still one of the most powerful and promising approaches to obtain the optical properties that go beyond those of their constituents. In this case the desired effects are achieved via subwavelength modulation of composite optical structures. Optical metamaterials are best known for negative refraction~\cite{smith2004metamaterials,shalaev2005negative,xiao2010loss}, hyperbolic light dispersion~\cite{poddubny2013hyperbolic,krokhin2016high}, artificial magnetism~\cite{monticone2014quest,o2002photonic,holloway2003double,wheeler2005three,popa2008compact,ginn2012realizing}, chirality \cite{fernandez2019new,ciattoni2015nonlocal,andryieuski2010homogenization,baranov2024effective} and other outstanding effects~\cite{cui2024roadmap}. Most of the mentioned effects were originally designed and demonstrated on plasmonic structures due to their negative permittivity in optical range. However, intrinsic Joule losses strongly limit their performance and prevent such structures from practical use.

In this context, in recent years there has been much greater attention to dielectric structures not subject to Joule losses. In order to achieve such non-natural effects as artificial magnetism it was proposed to employ the so-called Mie resonances~\cite{mie1908beitrage} that were observed experimentally in the whole visible spectral range from red to violet~\cite{kuznetsov2012magnetic,evlyukhin2012demonstration}. The study of Mie resonances opened a new direction of metamaterial physics~\cite{arbabi2015dielectric,kuznetsov2016optically,staude2017metamaterial,kivshar2018all}, which naturally raised an issue of applicability of effective medium approximation for corresponding resonant structures~\cite{simovski2018composite}.
It was shown that Mie resonances of the energy below the Bragg band get into the zone potentially suitable for homogenization and induce corresponding stop-band gaps~\cite{rybin2015phase}. In particular, it is well known that the first, magnetic-dipole-mode associated band gap is well described by a resonant pole of effective magnetic permeability, $\mu_\mathrm{eff}$,~\cite{o2002photonic,holloway2003double,wheeler2005three,popa2008compact,ginn2012realizing}.
Treatment of the second Mie resonance is commonly considered as much more complicated due to excitation of high-order multipole moments and strong non-locality.
Nevertheless, quadratic dispersion ($\omega\propto k^2$) of the eigenmodes at the bottom of the corresponding transmission band suggests that at least in some spectral range spatial dispersion effects might be accounted with reasonable efforts.
However, widely spread $\varepsilon_\mathrm{eff}-\mu_\mathrm{eff}$ approach is still fundamentally inapplicable, since effective permittivity and permeability describe only the density of dipole moments, but not densities of resonantly-enhanced high-order multipole moments, such as magnetic quadrupole one. In this scope, there is a high interest both in the development of theoretical model to advance in description of such metamaterials as well as to study the peculiar effects arising beyond the dipole approximation.

In this paper, we consider 2D photonic crystal structure of high-index cylinders. We experimentally vary the period of the structure and clearly demonstrate that at least the first two transverse electric (TE) Mie resonances get under the Bragg band in the region of affordable homogenization. Next, we apply an advanced homogenization procedure to go beyond the dipole approximation and describe second Mie resonance by effective medium approximation. For this purpose, we expand constitutive relations with specific magnetic quadrupole susceptibility and correspondingly update both dispersion relation and Fresnel equations.
We apply the developed effective medium approximation to calculate the spectrum of Fabry-Perot resonator made of quadrupolar metamaterial slab and validate calculations by finite element method.
In the considered case, contribution of magnetic quadrupole moment qualitatively changes the metamaterial behaviour by strong suppression of interface reflection.
The demonstrated results expand the opportunities to describe composite optical structures by sufficiently simple effective medium approximation and facilitate the design of optical devices made of the artificial materials.

\section{Results}

\subsection{Mie-resonant metamaterial}

\begin{figure}[h!]
    \centering
    \includegraphics[width=\linewidth]{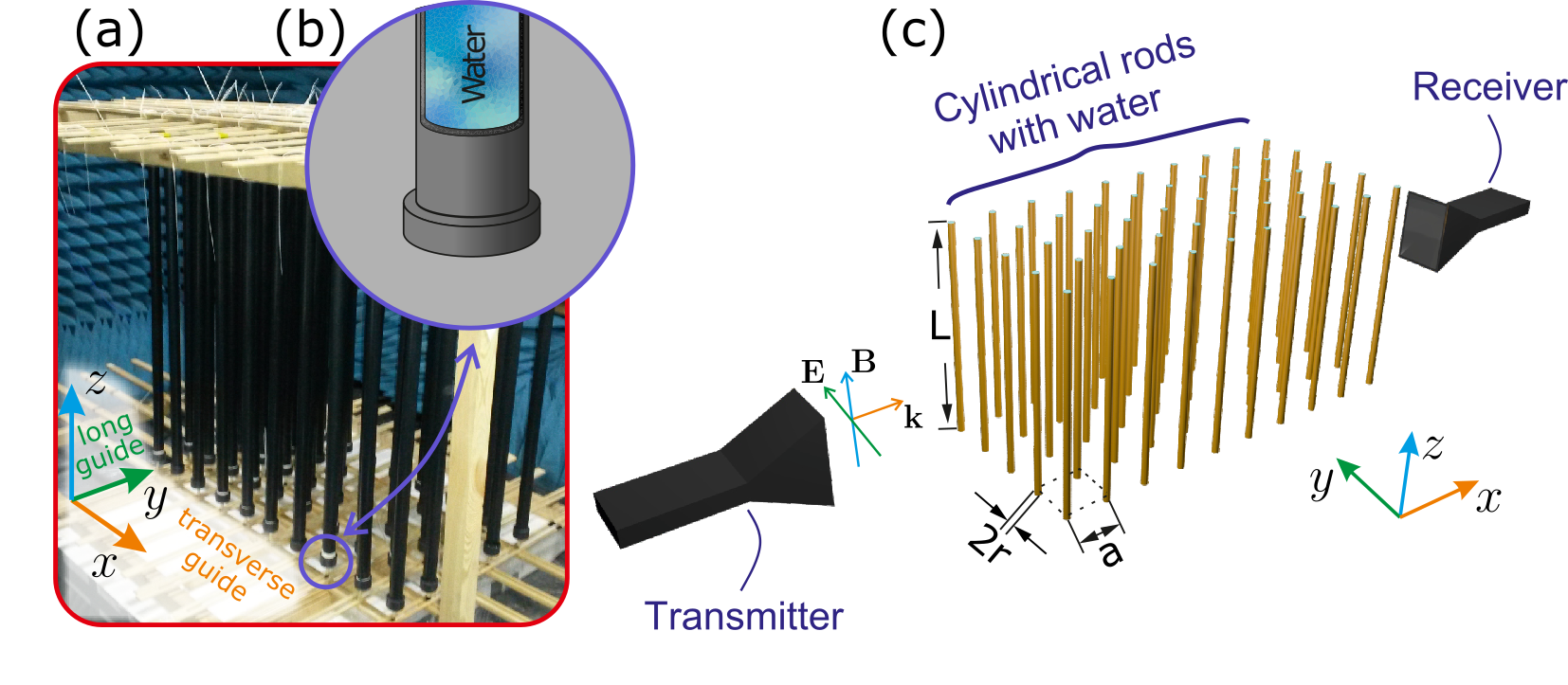}
    \caption{(a) Photo of a square grating of tubes filled with water. Period of the structure is tunable. (b) Schematic of the separate tube. (c) Schematic of the transmission measurements: structure is illuminated by TE-polarized emission from transmitter, which is in turn harvested by the receiver at the opposite side.}
    \label{fig:fig1}
\end{figure}

As an example of Mie-resonant metamaterial, we consider a square lattice of high-index cylindrical rods. Such structure is experimentally realized by 50 tubes of $r=13.6$~mm inner radius filled with deionized distilled water. A photograph of the experimental setup and its schematic are shown in Fig.~\ref{fig:fig1}~(a) and \ref{fig:fig1}~(b), respectively. The cylinders are made of polyvinyl chloride (PVC) tubes with sealed covers at the ends with a pair of thin pins on the upper and lower covers, which are used to fix the position of each cylinder in two types of guides along the $x$-axis and $y$-axis. The cylinders have an inner radius $r = 13.6$~mm, height $L = 1000$~mm and wall thickness of 2~mm. Structural elements of wood and PVC are practically invisible in the studied frequency range (0.5~GHz - 10.0~GHz) due to their low dielectric constant. Each pin is inserted into two sets of rails. 
The design provides an error of less than 2 mm in all directions and uniformity of movement for each row of cylinders. The metacrystal design allows a wide range of displacements along both axes. The ratio of the radius to the period lies in the range $r/a \approx$ 0.07–0.32.

Experimental studies are carried out in an anechoic chamber. To transmit and receive the signal a pair of TE-polarized TMA 1.0-18.0~GHz HF wideband horn antennas is used. Antennas are located at a distance of 1500~mm from the crystal boundaries (see Fig.~\ref{fig:fig1}~(c)), which corresponds to the far-field zone. The sample is irradiated with a TE-polarized field of $\lambda=30-400$~mm wavelength. Both antennas are connected to Agilent PNA E8362C microwave network analyzers operating in accumulation mode with an accumulation time of 5 seconds.

\begin{figure*}
    \centering
    \includegraphics[width=0.6\linewidth]{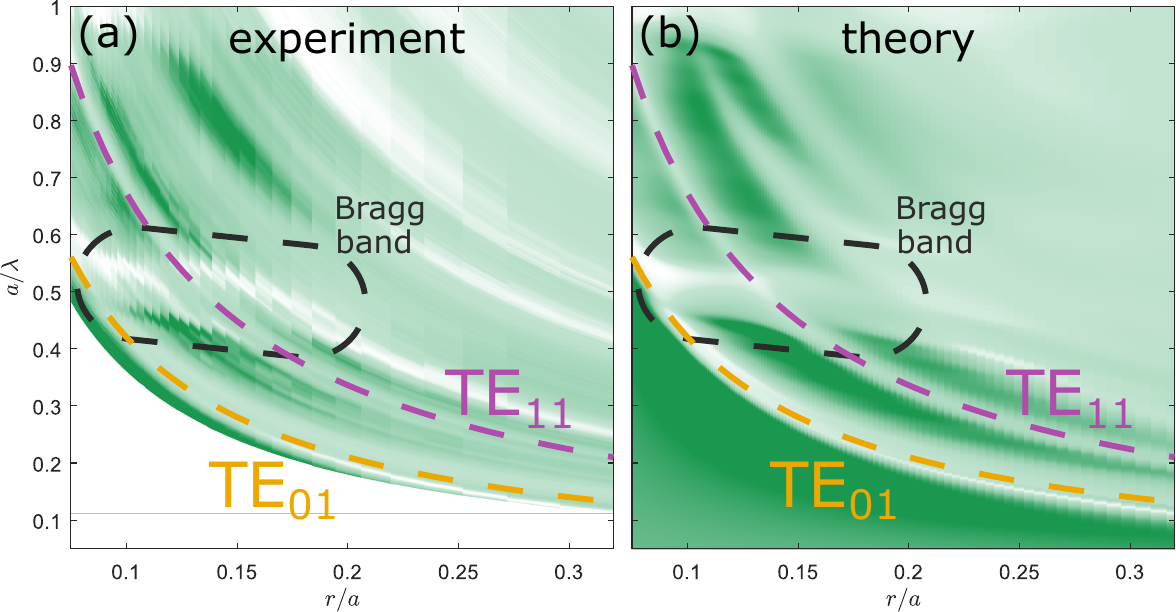}
    \caption{ Normalized transmission maps showing the spectrum dependence on the relative size of the tube, $r/a$, measured experimentally (a) and calculated numerically (b). Dashed black line indicates the Bragg stop-band. Yellow and magenta lines show the estimated energy of Mie modes of individual cylinders (for the sake of simplicity metaatoms' permittivity is assumed dispersionless $\varepsilon_\mathrm{cylinder}=80$, which makes the lines in the given axes hyperbolic).
    }
    \label{fig:fig2}
\end{figure*}

Experimentally measured and numerically calculated via multiple scattering theory spectral maps demonstrating normalized transmission spectrum dependence on the size parameter $r/a$ are shown in Fig.~\ref{fig:fig2}~(a) and~(b) respectively.
Spectral maps demonstrate a series of photonic transmission and stop bands. In particular, for the dimensionless energy in the range $a/\lambda \approx (0.45-0.55)$ there is a significant drop in transmission associated with Bragg reflection. Interestingly, Bragg band is almost insensitive for the filling factor of the lattice. Simultaneously, there is a number of stop-bands associated with Mie resonances of individual cylinders. Yellow and magenta dashed lines show analytical estimation for the first two TE-polarized Mie modes (TE${}_{01}$ and TE${}_{11}$, where the first index designates symmetry or orbital number of the mode and the second one enumerates the modes of the same symmetry) of individual cylinders. For our estimation we assume permittivity of water frequency independent, $\varepsilon_\mathrm{H_2O}=80$, and do not account for neighbors interactions, which makes the shape of dashed lines hyperbolic. Nevertheless, even such rough estimation matches corresponding associated stop-bands of the grating both for experimental and numerical results.

Experimental spectra represent rather good correspondence with theoretical data, which is clearly seen from spectral maps (Fig.~\ref{fig:fig2}~(a-b)).
But what is way more important, these results convincingly demonstrate that we have experimentally achieved decent of TE${}_{01}$ and TE${}_{11}$-generated stop-bands below the Bragg gap. In other words, we have achieved the conditions for which the structure might be potentially considered as Mie-resonant metamaterial. Moreover, the second mode is associated with resonant excitation of magnetic quadrupole moment, which makes introduction of appropriate effective medium approximation challenging.

\subsection{Homogenization of Quadrupole Metamaterial}

In order to describe quadrupole metamaterials by some macroscopic theoretical model, we obviously need to consider density of magnetic quadrupole moment $\hat{S}$ along with densities of electric, $\mathbf{P}$, and magnetic, $\mathbf{M}$, dipole moments (see Appendix~\ref{app:multipole_definition} for definitions). In this scope, we introduce the following model connecting multipolar densities with in-plane components of macroscopic electric fields $\mathbf{E}_\parallel = \left(E_x,E_y\right)^\mathrm{T}$ and its gradients:

\begin{align}
    \begin{pmatrix}
        {P}_{x}\\
        {P}_{y}
    \end{pmatrix} & =\chi^{P}_0         \begin{pmatrix}
        {E}_{x}\\
        {E}_{y}
    \end{pmatrix}+\frac{\chi^{P}_2}{k_0^2}\left(\nabla_\parallel\otimes\nabla_\parallel-\triangle_\parallel\right)
        \begin{pmatrix}
        {E}_{x}\\
        {E}_{y}
    \end{pmatrix},\label{eq:sus1}\\
    {M}_z& = -i\chi^M_1\left[\frac{\nabla_\parallel}{k_0}\times\mathbf{E}_\parallel\right]_z=\chi^M_1{B}_z,\label{eq:sus2}\\
    \begin{pmatrix}
        {S}_{zx}\\
        {S}_{zy}
    \end{pmatrix} &=-\frac{i}{k_0}\chi^{S}_0\left[
        \mathbf{z}\times\mathbf{E}_\parallel\right]_\parallel,\label{eq:sus3}
\end{align}
where $\chi_0^P$, $\chi_2^P$, $\chi_1^M$ and $\chi_0^S$ are multipolar susceptibilities; $k_0=\omega/c$ is light wavevector in vacuum, differential operators are defined as follows $\nabla_\parallel = \left(\partial_x,\partial_y\right)^\mathrm{T}$, $\triangle_\parallel = \partial^2_x+\partial^2_y$ and $\otimes$ denotes tensor product.
In fact, this model is Tailor-like expansion of multipole polarization densities over macroscopic electric field with only most important terms left. For the magnetic quadrupole and magnetic dipole terms we account only for the first non-zero terms, which are proportional to electric field and its gradient correspondingly. Moreover, due to the symmetry of the structure, we consider the response of the magnetization, $M$, only of the anti-symmetric combination of the gradients, which is effectively curl of the electric field, which is in turn proportional to magnetic field. For the electric dipole polarization except from the conventional term proportional to electric field we also account for the response on the second gradient of it. Moreover, we consider only such contributions that preserve circular shape of isofrequency contours. Although this term might seem non-essential and excessive, it will be further shown that it provides spatial dispersion corrections to the currents excited in the structure of the same order as magnetic dipole and quadrupole moments.

\begin{figure*}
    \centering
    \includegraphics[width=0.95\linewidth]{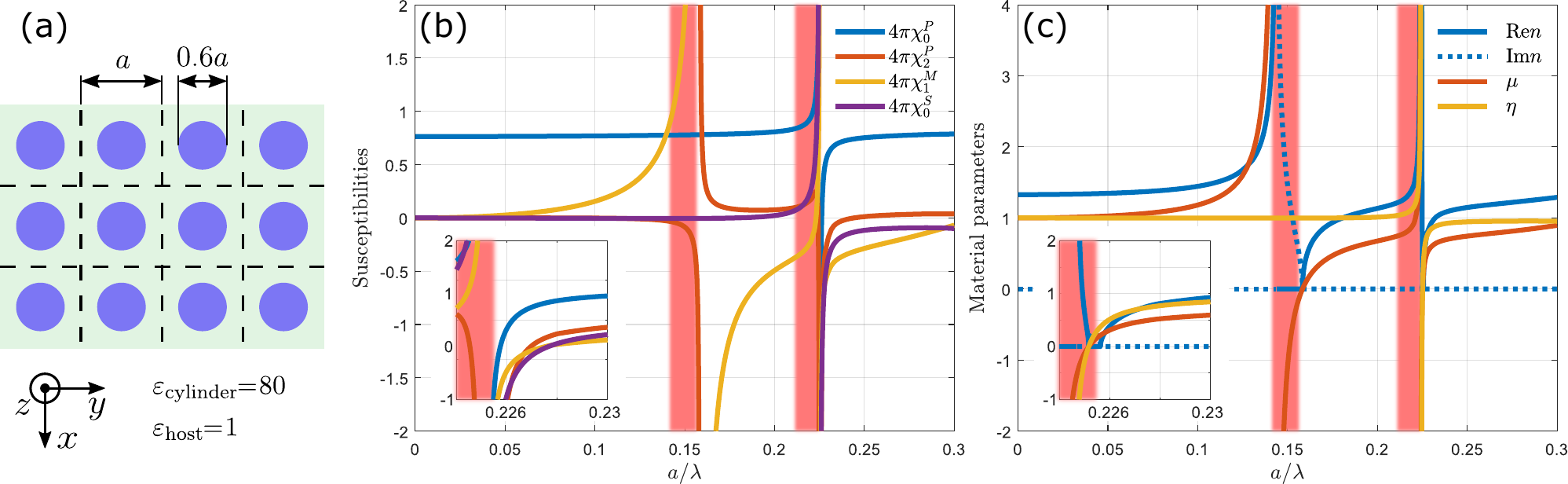}
    \caption{(a) Schematic of the structure considered for homogenization. The square lattice consists of high-permittivity cylinders $\varepsilon_\mathrm{cylider}=80$ of $r=0.3a$ in a "void" host medium $\varepsilon_\mathrm{host}=1$. (b) Spectra of multipolar susceptibilities of the considered metameterial. In the vicinity of the first Mie resonant stop-band ($a/\lambda\approx 0.15$), the electric dipole, $\chi_0^P$, and magnetic quadrupole, $\chi_0^S$, susceptibilities remain continuous and almost constant, whereas the second stop-band ($a/\lambda\approx 0.22$) results in divergence of all the considered susceptibilities.  (c) Spectra of the material parameters of the metamaterial, which are convenient both for the description of the dispersion of the eigenmodes and their boundary conditions.  
    Pink stripes quantitatively indicate the bands of expected poor performance of the homogenization models.}
    \label{fig:fig3}
\end{figure*}

We extract all four required susceptibilities from microscopic calculations conducted by the finite element method (FEM) in COMSOL Multyphysics. The adaptation procedure of the current-injection-based approach~\cite{silveirinha2007metamaterial} for specific purposes of our study is discussed in detail in Appendix. In order to simplify the analysis, we consider water as dispersionless material of $\varepsilon_{\mathrm{H_2O}}=80$ permittivity. It enables scaling and allows to consider structures of arbitrary size without loss of generality. Moreover, for the radio band such approximation is rather good, whereas the behaviour at other frequencies is out of our interest. Radius of the structure is taken for $r=0.3a$ (see Fig.~\ref{fig:fig3}~(a) for the schematic of the structure), which is close to the upper limit available in experiment (see Fig.~\ref{fig:fig2}~(a)) and provides Mie resonances deep below the the Bragg band.

Spectra of the calculated susceptibilities are shown in Fig.~\ref{fig:fig3}~(b). We observe that the first Mie resonance at $a/\lambda\approx0.15$ is primarily associated with the divergence of magnetic dipole susceptibility, $\chi_1^{M}$, and substantially more suppressed resonance of $\chi_2^{P}$ responsible for the correction of electric dipole polarization. Notably, both magnetic quadrupole, $\chi_0^{S}$, and electric dipole, $\chi_2^{P}$, susceptibilities are absolutely insensitive to the effect. For the second Mie resonance at $a/\lambda \approx 0.22$ all the susceptibilities experience divergence, which makes the behaviour much more complicated. Indeed, the modes of $D_{\infty h}$-symmetric square grating of cylinders have different symmetry from the Cartesian multipoles and, therefore, some resonance might easily excite a number of different multipolar moments. Pink stripes quantitatively indicate the bands of expected poor performance of the homogenization model. They are slightly red-shifted relative to the stop-bands and are associated with large values of either real or imaginary parts of $\mathbf{k}$-vector of eigenmodes. The first case is observed right below the stop-band, where dispersion is close to the boundary of the Brillouin zone. The second one is observed for the most part of the stop-band (except for the very top), where the penetration depth is of order of a single unit cell.

Based on these four susceptibilities it is possible to formulate both the dispersion relation and boundary conditions. Indeed, the dispersion relation is a solution of the following equation:
\begin{equation}
    \left[\triangle-\nabla\otimes\nabla+k_0^2\right]\mathbf{E}=\frac{-4\pi i k_0}{c}\mathbf{j}=-4\pi  k^2_0\mathbf{P}^\mathrm{gen}\left[\mathbf{E}\right],
\end{equation}
where $\mathbf{P}^\mathrm{gen}=\frac{\mathbf{j}}{-i\omega}$ is the generalized macroscopic polarization associated with the total polarization current in medium, $\mathbf{j}$. This polarization can be found as a sum of electric dipole polarization, curl of magnetization and gradients of high multipole moments densities~\cite{raab2005multipole,simovski2018composite,evlyukhin2016optical,silveirinha2007metamaterial,jackson1998classical}:
\begin{equation}
    \mathbf{P}^\mathrm{gen}=\mathbf{P}+\frac{1}{-ik_0}\nabla\times\mathbf{M}-\frac{1}{2}\nabla\cdot\hat{\mathbf{Q}}-\frac{1}{-2ik_0}\nabla\times\hat{\mathbf{S}}\cdot\nabla+\dots\label{eq:eq5},
\end{equation}
which in our particular case is reduced to:
\begin{multline}
    \begin{pmatrix}{P}^\mathrm{gen}_x\\{P}^\mathrm{gen}_y
    \end{pmatrix}=
     \chi_0^P \begin{pmatrix}{E}_x\\{E}_y
    \end{pmatrix} -\\-\frac{\chi_2^P+    \chi_1^M+ \chi_0^S/2}{k_0^2} \begin{pmatrix}
        \partial_y^2 & -\partial_x\partial_y\\
        -\partial_x\partial_y & \partial_x^2
    \end{pmatrix}\begin{pmatrix}{E}_x\\{E}_y
    \end{pmatrix}.
\end{multline}
As we can see three of the susceptibilities (namely $\chi_2^P$, $\chi_1^M$, $\chi_0^S$) are included in the expressions as a common coefficient responsible for the non-local contribution to the polarization current induced by the second derivatives of electric field. Gathering them all together, we finally obtain a self-consistent equation for eigenmodes in the bulk:

\begin{multline}
    \left(-1+4\pi \left(\chi_2^P+    \chi_1^M+ \chi_0^S/2\right)\right)\begin{pmatrix}
        \partial_y^2 & -\partial_x\partial_y\\
        -\partial_x\partial_y & \partial_x^2
    \end{pmatrix}\begin{pmatrix}{E}_x\\{E}_y
    \end{pmatrix}\\=
    \left(1+4\pi   \chi_0^P\right)k_0^2 \begin{pmatrix}{E}_x\\{E}_y
    \end{pmatrix}.
\end{multline}

Although one can explore this equation for the solutions of arbitrary polarization, we will be focused on orthogonal waves
($\mathbf{E}_\parallel\perp\mathbf{k}_\parallel$).
Assuming $\left({E}_x, {E}_y\right)^\mathrm{T}\propto \left(-k_y,k_x\right)^\mathrm{T}e^{i\mathbf{kr}}$, we finally obtain the dispersion law, which allows us to introduce the refractive index $n(\omega)=k/k_0$ as a first material parameter:
\begin{equation}
    k^2=n^2k_0^2 = \frac{1+4\pi\chi_0^P}{1-4\pi\left(\chi_2^P+\chi_1^M+ \chi_0^S/2\right)}k_0^2.\label{eq:dispersion}
\end{equation}

However, not only the dispersion of the bulk waves determines the behaviour of the metamaterial. We are also interested in appropriate boundary conditions~\cite{silveirinha2014boundary,graham2000multipole,raab2005multipole,gorlach2020boundary}. In order to obtain them, we apply theoretical results from~\cite{graham2000multipole} for our particular model (see Appendix~\ref{sec:boundary_conditions}-\ref{sec:fresnel} for details). For the considered polarization and boundary $x=\mathrm{const}$ we have the following conditions:
\begin{align}
    B_z\left[1-4\pi\left(\chi_1^M+\chi_0^S/2\right)\right]=\mathrm{const} \label{eq:BC_magnetic}\\
    E_y(1-2\pi\chi_0^S)=\mathrm{const} \label{eq:BC_electric}
\end{align}
As we can see, susceptibilities of the magnetic dipole and quadrupole moments are included in dispersion relation (Eq.~\ref{eq:dispersion}) and boundary condition for magnetic field (Eq.~\ref{eq:BC_magnetic}) in the same combination $\chi_1^M+\chi_0^S/2$. At first sight, it might seem that the quadrupole susceptibility might be considered just as a correction to the dipole one. Nevertheless, it is quadrupole susceptibility that solely determines boundary condition for electric field (Eq.~\ref{eq:BC_electric}) and, moreover, makes the transverse components of electric field discontinuous.
All other physical effects except this one could be attributed by appropriate corrections to $\varepsilon_\mathrm{eff}$ and $\mu_\mathrm{eff}$ parameters.

\begin{figure*}
    \centering
    \includegraphics[width=0.6\linewidth]{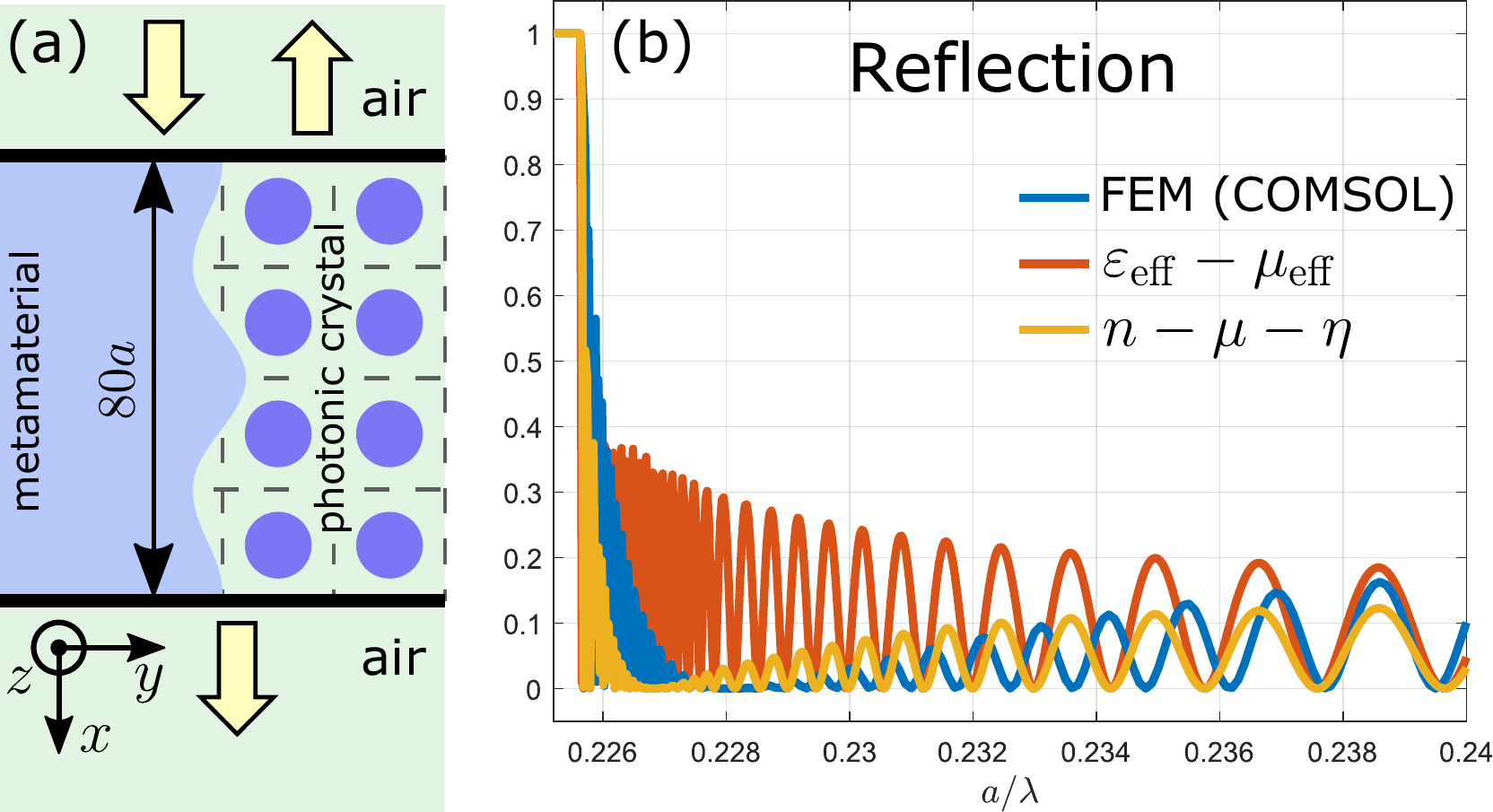}
    \caption{(a) Schematic of 80-layers Fabry-Perot slab under normal incidence of light. It might be considered both as photonic crystal and as homogenized metamaterial. (b) Reflection spectra for the energies right above the second Mie-resonant band gap. The reference, FEM-calculated spectrum is compared with spectra obtained via different homogenization approaches. The conventional $\varepsilon_\mathrm{eff}-\mu_\mathrm{eff}$ approaches generally well describes the opening of the transmission band and period of the oscillations. At the same time, accounting for the quadrupole magnetization by the developed $n-\mu-\eta$ approach allows to reproduce envelope of the oscillations as well.
    }
    \label{fig:fig4}
\end{figure*}

Finally, putting it all together, we introduce the following definitions for three material parameters:
\begin{align}
    n (\omega)& = \sqrt{\frac{1+4\pi\chi_0^P}{1-4\pi\left(\chi_2^P+\chi_1^M+ \chi_0^S/2\right)}},\label{eq:n}\\
    \mu (\omega) & = \frac{1}{1-4\pi\left(\chi_1^M+ \chi_0^S/2\right)},\\
    \eta (\omega)& = \frac{1}{1-2\pi \chi_0^S},
\end{align}
where $n$ and $\mu$ are introduced in such a way that they match the definition of refractive index and magnetic permeability in the absence of magnetic quadrupole moment. Concurrently, $\eta$, represents some kind of a magnetic quadrupole permeability that enables convenient description of corresponding boundary condition. In Fig.~\ref{fig:fig3}~(c) we demonstrate spectra of our just introduced material parameters. 
From Fig.~\ref{fig:fig3}~(c) one can see that metamaterial's
effective refractive index is purely real for transmission bands and purely imaginary for the~stop bands induced by Mie resonances, since all the susceptibilities are purely real in our case and according to Eq.~\ref{eq:dispersion}, $k^2/k_0^2$, is real as well. Magnetic dipole and quadrupole permeabilities are purely real for the whole spectrum as well. Remarkably, the first Mie resonance results in a divergence of only magnetic dipole permeability, $\mu$, but not quadrupole one, $\eta$, which does not "feel" the resonance at all. At the same time, the second mode affects all the materials parameters of our scope, which is in accordance with the physical processes occurring.

Having these three parameters, we can formulate the dispersion relation and boundary conditions in a concise form:
\begin{align}
     k_\parallel & =  n (\omega)k_0,\\
    \frac{B_z}{\mu (\omega)} & =\mathrm{const},\\
    \frac{E_y}{\eta (\omega)} & = \mathrm{const}.
\end{align}

Together, they fully describe light propagation in quadrupole metamaterials, including interfaces between them. In particular, we apply the obtained boundary conditions to revise classical Fresnel reflection problem for the case of quadrupole metamaterials. For TE-polarization of our interest, we derive the following Fresnel coefficients (see Appendix~\ref{sec:fresnel}-\ref{sec:reflection} for derivations and illustrations):

\begin{align}
     t &= \frac{2\mu_1n^2_2\eta_2 k_{x1}}{\mu_1n^2_2\eta_2 k_{x1}+\mu_2n^2_1\eta_1 k_{x2}}\frac{\mu_2 n_1}{\mu_1 n_2},\\
    r &= \frac{\mu_1n^2_2\eta_2 k_{x1}-\mu_2n^2_1\eta_1 k_{x2}}{\mu_1n^2_2\eta_2 k_{x1}+\mu_2n^2_1\eta_1 k_{x2}}.
\end{align}

where $k_{x,1(2)} = \sqrt{n_{1(2)}^2k_0^2-k_y^2}$ is the value of the wavevector component perpendicular to the interface and indices 1 and 2 specify incoming and outgoing media from which the wave falls and to which it is transmitted correspondingly.

In turn, based on the Fresnel equations we are able to consider any layered structures of quadrupole metamaterials. In order to verify our theoretical model, we consider an 80-period-thick slab of metamaterial in vacuum as shown in Fig.~\ref{fig:fig4} and calculate the corresponding reflection spectrum via different approaches. We are mostly interested in a frequency band for which magnetic quadrupole contribution is important, but we still can neglect all the other high multipolar contributions. As it has been already mentioned, these conditions are expected at the opening of the transmission band above the second Mie resonance, which is considered below.

The reference spectrum calculated for the true photonic crystal structure via the FEM in COMSOL Multiphysics is depicted by blue solid lines (see Fig.~\ref{fig:fig4}~(b-c)). When we try to calculate this spectra in the framework of conventional $\varepsilon_\mathrm{eff} - \mu_\mathrm{eff}$ effective medium approximation~\cite{landau2013electrodynamics} (see Fig~\ref{fig:fig4}~(b)) that accounts only for the dipole polarizations, we observe that such metamaterial approximation fails to describe the shape of the envelop function of Fabry-Perot oscillations. At the same time, since the period of the oscillations is generally similar, the most crucial deviations of such model occur not for the refractive index, but for the boundary conditions. In this way, there is no surprise, that when we take into account a magnetic quadrupole moment, resulting spectrum becomes much closer to the reference one (see Fig.~\ref{fig:fig4}~(b)). Now the main behavior is reproduced accurately. Right above the stop-band the reflection tends to zero and only then slowly ascends again. Although, we observe a slight shift in the frequency of zero-reflection point and a slight mismatch in the phase of oscillations, the general behavior is reproduced correctly. It is worth noting separately, that the envelop of Fabry-Perot oscillations reaches zero point at some energy. This implies that at this point reflection is fully suppressed not because of destructive interference in the slab, but due to the absence of reflection from each of the interfaces. In other words, we observe some sort of the Brewster effect for the normal incidence.
Validation of our model for the case of Fabry-Perot slab demonstrates the potential to accurately describe other, much more complicated macroscopic structures of magnetic quadrupole metamaterials.

\section{Conclusions}

In this paper, we have experimentally demonstrated the emergence of magnetic quadrupole metamaterial. We have developed a theoretical model for its homogenization and demonstrated that the quadrupole polarization leads to discontinuities of the tangential component of macroscopic electric field at the interface. Associated correction of boundary conditions might result in zero reflection for normal incidence and reconfiguration of reflection/transmission spectra in general. The obtained results are successfully validated by independent FEM calculations for Fabry-Perot metamaterial slab.

\section{Acknowledgements}

The work of D.G.B., M.V.R., and S.A.D. was supported by the ITMO-MIPT-Skoltech Clover Program.
The work of D.S.F. and A.V.N. was supported by the Ministry of Science and Higher Education of the Russian Federation (FSMG-2025-0005).

A.V.N., D.S.F., K.B.S. and M.F.L. fabricated experimental setup and conducted measurements.
M.V.R. calculated transmission spectra of experimental setup.
I.M.F., N.S.S., S.A.D. and N.A.G. developed homogenization approach and conducted corresponding calculations.
I.M.F. prepared the manuscript.
M.V.R., K.B.S., M.F.L., and
N.A.G. supervised experimental and theoretical parts of the study.
All authors reviewed and edited the paper. All authors contributed to the discussions and commented on the paper.

\appendix

\section{On definition of multipole moments}

\label{app:multipole_definition}

There are several different definitions of multipole moments that can be found in the literature. First, misunderstanding might be caused by confusion of Cartesian and so-called exact~\cite{alaee2018electromagnetic,alaee2019exact} multipole moments. The latter ones are best for the description of currents localized in individual particles, whereas the distributed currents in metamaterials are conventionally described by densities of Cartesian moments.

Second, there are discrepancies even in the definition of Cartesian multipoles associated with their symmetrization and normalization. Indeed, different multipole moments correspond to tensors of different ranks, but these tensors can be reduced to irreducible representation, reduced partially or not reduced at all. For instance, magnetic dipole moment emerges from electric quadrupole tensor as its antisymmetric part. In the same way, magnetic quadrupole moment results from symmetrization of electric octupole moment. Full reduction of magnetic quadrupole-electric octupole tensor results in manifestation of widely-known toroidal dipole moment~\cite{evlyukhin2016optical} as well. For the sake of convenience, in this paper we symmetrize electric quadrupole and octupole moments, but not the magnetic quadrupole one. The reason is that our fundamentally 2D problem requires consideration of only $S_{zx}$ and $S_{zy}$ components of tensor and there is no reason to complicate all the derivations by mixing them with components $S_{xz}$ and $S_{yz}$ possessing drastically different optical properties. In this way, we employ the following definitions of multipole moments:

\begin{align}
    \mathbf{p} &=\frac{1}{-i\omega}\int_V\mathbf{j}(\mathbf{r})d^3\mathbf{r},\\
    \mathbf{m} &=\frac{1}{2c}\int_V\mathbf{r}\times\mathbf{j}(\mathbf{r})d^3\mathbf{r},\label{eq:A2}\\
    \hat{\mathbf{q}} &=\frac{1}{-i\omega}\int_V\left[\mathbf{r}\otimes\mathbf{j}(\mathbf{r})+\mathbf{j}(\mathbf{r})\otimes\mathbf{r}\right]d^3\mathbf{r},\\
    \hat{\mathbf{s}} &=\frac{2}{3c}\int_V\left[\mathbf{r}\times\mathbf{j}(\mathbf{r})\right]\otimes\mathbf{r}d^3\mathbf{r}.
\end{align}

\section{Theoretical background of homogenization}

Homogenization of the composite structures such as metamaterials is a procedure, which allows to "average" microscopic electromagnetic fields and polarization currents inside some composite structure to deal with macroscopic fields. It allows one to get rid of consideration of subwavelength peculiarities and phase oscillations. Such approach strongly simplifies numerical calculations and explains the physical processes occurring in much more intuitive way. In practice, most of the studies are limited by taking into account macroscopic densities of only electric, $\mathbf{P}$, and magnetic, $\mathbf{M}$, dipole moments and their response on macroscopic electric and magnetic fields~\footnote{An apparent optical response on macroscopic magnetic field is in fact associated with non-local response on electric field. However, in electrodynamics corresponding effects are indistinguishable.}. High multipole moments as well as high order non-local effects of their optical response are conventionally ignored as they complicate the homogenization procedure so much that it does not pay off. For this reason, corresponding optical effects are attributed to photonic crystal regime and are considered within appropriate numerical approaches. However, some contributions of the high multipoles provide the non-locality of the same order as well-known artificial magnetism.

In order to clarify this idea, we need to recap the basics of the homogenization procedure. The main idea of the approach is that we would like to substitute the true Maxwell equations describing microscopic electromagnetic fields and currents with Maxwell equations operating smooth, slowly varying in space macroscopic fields. For the sake of brevity we consider only the last Maxwell equation:

\begin{equation}
\begin{gathered}
   \nabla\times\mathbf{B}^\mathrm{micro}=\frac{4\pi}{c}\mathbf{j}^\mathrm{micro}-ik_0\mathbf{E}^\mathrm{micro}\\
   \ \downarrow\ \downarrow\ \downarrow\ \downarrow\ \downarrow\ \downarrow\ \downarrow\ \downarrow\ \downarrow\ \downarrow   \\
   \nabla\times\mathbf{B}^\mathrm{macro}=\frac{4\pi}{c}\mathbf{j}^\mathrm{macro}-ik_0\mathbf{E}^\mathrm{macro} 
\end{gathered}\label{eq:transform}
\end{equation}

In order to obtain macroscopic values of the fields we need to average microscopic ones over such small volume that phase oscillations are negligible. In other words, typical dimensions of the averaging domain should be much smaller than wavelength in artificial medium. In the long wavelength limit we can assume that wavelength is much larger than the period of the structure, $\lambda_{\mathrm{MM}}\gg a$, and utilize the unit cell as an averaging or integration domain. Unfortunately, this condition leaves a very narrow range of applicability and does not allow consideration of most of the effects of interest related to the spatial dispersion. For this reason, we apply a dynamic averaging procedure~\cite{simovski2018composite}. In this case we consider a plane wave solution determined by the wavevector, $\mathbf{k}$. Macroscopic fields are expected to oscillate as follows:

\begin{align}
    \mathbf{E}^\mathrm{macro}  &=  \tilde{\mathbf{E}}^\mathrm{macro}e^{i\mathbf{kr}}\\
    \mathbf{B}^\mathrm{macro}  &=  \tilde{\mathbf{B}}^\mathrm{macro}e^{i\mathbf{kr}}\\
    \mathbf{j}^\mathrm{macro}  &=  \tilde{\mathbf{j}}^\mathrm{macro}e^{i\mathbf{kr}},
\end{align}
where $\tilde{\mathbf{E}}^\mathrm{macro}$, $\tilde{\mathbf{B}}^\mathrm{macro}$ are the amplitudes of macroscopic electric and magnetic fields and $\tilde{\mathbf{j}}^\mathrm{macro}$ is an amplitude of the macroscopic polarization current. One might guess that these amplitudes can be found via averaging over a unit cell with accounting for "retarded" phase, which is known as a dynamic averaging procedure:

\begin{align}
    \tilde{\mathbf{E}}^\mathrm{macro}&=\frac{e^{i\mathbf{kr}}}{V}\int_V\mathbf{E}^\mathrm{micro}(\mathbf{r}')e^{-i\mathbf{kr'}}d^3\mathbf{r}',\\
    \tilde{\mathbf{B}}^\mathrm{macro}&=\frac{e^{i\mathbf{kr}}}{V}\int_V\mathbf{B}^\mathrm{micro}(\mathbf{r}')e^{-i\mathbf{kr'}}d^3\mathbf{r}',\\
    \tilde{\mathbf{j}}^\mathrm{macro}&=\frac{e^{i\mathbf{kr}}}{V}\int_V\mathbf{j}^\mathrm{micro}(\mathbf{r}')e^{-i\mathbf{kr'}}d^3\mathbf{r}'.
\end{align}
It might be easily shown that such procedure allows to formulate Maxwell equation for macroscopic fields in the original form (see Eq.~\ref{eq:transform}). However, such averaging is just the first step in the process of homogenization. We would like to determine the connection of polarization current and electromagnetic fields accounting for both temporal and spatial dispersion. Therefore, we need to find some microscopic fields to average for any arbitrary combination of $\omega$ and $\mathbf{k}$, which is not a trivial problem when $\omega$ and $\mathbf{k}$ do not correspond to eigenmodes of the structure. There are different approaches to tackle this issue~\cite{ciattoni2015nonlocal,andryieuski2010homogenization,baranov2024effective,gorlach2020boundary,menzel2008retrieving,repan2021artificial,petschulat2008multipole,gorlach2014effect,simovski2007local,simovski2010electromagnetic,alu2011first}, but we follow the original study by Silveirinha~\cite{silveirinha2007metamaterial}, where it is suggested to consider not eigenmodes of the structure, but electromagnetic waves excited by some external current $\mathbf{j}^\mathrm{ext}(\omega,\mathbf{k}) \propto e^{i\mathbf{kr}}e^{-i\omega t}$ at the desired temporal and spatial frequencies.

Now, being able to calculate macroscopic polarizations of artificial material and macroscopic electric field we are in a position to formulate constitutive relations. In most cases it is more convenient to operate not with the calculated macroscopic current but with corresponding macroscopic generalized polarization $\mathbf{P}^\mathrm{gen}(\omega,\mathbf{k})=\frac{\mathbf{j}^\mathrm{macro}(\omega,\mathbf{k})}{-i\omega}$. If we are only interested in the dispersion of the eigenmodes in the bulk~\cite{landau2013electrodynamics}, then we might introduce generalized permittivity tensor $\hat{\varepsilon}^\mathrm{gen}$:

\begin{equation}
    \hat{\varepsilon}^\mathrm{gen}(\omega,\mathbf{k}) \mathbf{E}^{\mathrm{macro}}(\omega,\mathbf{k})=\mathbf{E}^{\mathrm{macro}}(\omega,\mathbf{k})+4\pi\mathbf{P}^{\mathrm{gen}}(\omega,\mathbf{k}),\label{eq:0}
\end{equation}
which is the only property of the medium, which enters the dispersion relation~\cite{agranovich2013crystal,landau2013electrodynamics}:

\begin{equation}
    \left(k^2-\mathbf{k}\otimes\mathbf{k}\right)\mathbf{E}=\hat{\varepsilon}^\mathrm{gen}k_0^2\mathbf{E}.
\end{equation}

Unfortunately, the dispersion of the modes does not fully describe optical properties of material. In particular, knowledge of the dispersion does not guarantee the possibility of formulation of appropriate boundary conditions~\cite{silveirinha2014boundary,graham2000multipole,raab2005multipole,gorlach2020boundary}. The reason for that is encoded inside the generalized polarization, which is associated not only with a contribution of electric dipole moment density, but also with the curl of magnetization as well as high gradients of all the other multipole moments~\cite{raab2005multipole,simovski2018composite,evlyukhin2016optical,silveirinha2007metamaterial,jackson1998classical}:

\begin{equation}
    \mathbf{P}^\mathrm{gen}=\mathbf{P}+\frac{1}{-ik_0}\nabla\times\mathbf{M}-\frac{1}{2}\nabla\cdot\hat{\mathbf{Q}}-\frac{1}{-2ik_0}\nabla\times\hat{\mathbf{S}}\cdot\nabla+\dots, \label{eq:1}
\end{equation}
where $\mathbf{P}$, $\mathbf{M}$, are macroscopic densities of electric and magnetic dipole moments and $\hat{\mathbf{Q}}$, $\hat{\mathbf{S}}$  are densities of electric and magnetic quadrupole moments.
All the terms in sum~\ref{eq:1} are associated with currents indistinguishable in the bulk of the structure and only their cumulative value $\mathbf{P}^\mathrm{gen}$ determines the bulk dispersion. However, if we consider an interface of artificial material, each multipole moment will provide its own specific contribution to the surface current. Obviously, these surface currents result in different boundary conditions~\cite{silveirinha2014boundary,graham2000multipole,raab2005multipole,gorlach2020boundary}.

For this reason, in order to fully describe metamaterial by effective medium approximation we need to consider all the multipole moments from~Eq.~\ref{eq:1} and formulate corresponding constitutive relations. In principle, we might connect multipole moments densities with only macroscopic electric field as it was done for generalized polarizability in Eq.~\ref{eq:0}, but in this case the response would be non-local and we would still need to tackle spatial dispersion effects. Straightforward approach to reformulate problem in a local way is to consider the response on not only electric field, but on its gradients as well. Constitutive relations might be formulated as follows:

\begin{equation}
    \begin{pmatrix}
        \mathbf{P}\\
        \mathbf{M}\\
        \hat{\mathbf{Q}}\\
        \hat{\mathbf{S}}\\
        \dots
    \end{pmatrix} = \begin{pmatrix}
        \colorbox{light_green}{$\hat{\chi}^{P}_0$}         &        \colorbox{light_blue}{$\hat{\chi}^{P}_1$}    & \colorbox{light_orange}{$\hat{\chi}^{P}_2$} &\dots\\
        \colorbox{light_blue}{$\hat{\chi}^{M}_0$}         & \colorbox{light_orange}{$\hat{\chi}^{M}_1$}   & \hat{\chi}^{M}_2 &\dots\\
        \colorbox{light_blue}{$\hat{\chi}^{Q}_0$}         & \colorbox{light_orange}{$\hat{\chi}^{Q}_1$}   & \hat{\chi}^{Q}_2 &\dots\\
        \colorbox{light_orange}{$\hat{\chi}^{S}_0$}         & \hat{\chi}^{S}_1   & \hat{\chi}^{S}_2 &\dots\\
        \cdots             &\dots        &\dots      &\dots
        \end{pmatrix}
        \begin{pmatrix}
        \mathbf{E}\\
        \nabla\otimes\mathbf{E}\\
        \nabla\otimes\nabla\otimes\mathbf{E}\\
        \dots
    \end{pmatrix}, \label{eq:2}
\end{equation}
where $\hat{\chi}$ are multipolar susceptibility tensors,  $\left(\nabla\otimes\mathbf{E}\right)_{ij}=\partial_iE_j$ and $\left(\nabla\otimes\nabla\otimes\mathbf{E}\right)_{ijk}=\partial_i\partial_jE_k$ are tensors of partial derivatives.
    
If we substitute Eq.~\ref{eq:2} into Eq.~\ref{eq:1} we will obtain generalized polarization, $\mathbf{P}^\mathrm{gen}$, as a series over the derivatives of electric field:

\begin{equation}
        \mathbf{P}^\mathrm{gen}=\colorbox{light_green}{$\hat{a}(\omega)$}\mathbf{E}+\colorbox{light_blue}{$\hat{b}(\omega)$}\nabla\otimes\mathbf{E}+\colorbox{light_orange}{$\hat{c}(\omega)$}\nabla\otimes\nabla\otimes\mathbf{E}+\dots, \label{eq:3}
\end{equation}
It can be easily noticed that susceptibilities highlighted by specific colors correspond to the contributions of the same order. In particular, the light-green colored susceptibility is the only one, which contributes to the term proportional to electric field $\mathbf{E}$ ($\hat{a}(\omega)$ in~Eq.~\ref{eq:3}), the light-blue ones correspond to the term of the first gradient of the electric field ($\hat{b}(\omega)$ in~Eq.~\ref{eq:3}) and the orange ones are responsible for the derivative of the second order ($\hat{c}(\omega)$ in~Eq.~\ref{eq:3}).

In practice, we can not consider the infinite series of multipole moments and electric field gradients and truncate them. Nevertheless, even when we consider the approximate model, it is highly desirable to account all the susceptibilities of the same order, otherwise the translation invariance can be broken~\cite{simovski2018composite}.

In our particular study, the structure possesses an inversion center and, therefore, linear contributions to spatial dispersion are prohibited, i.e. the light-blue terms are nullified. Among the other colored terms, we account for the main components of $\hat{\chi}_0^P$, $\hat{\chi}_0^S$, but neglect $\chi_1^Q$ as a small one. Regarding $\chi_1^M$ and $\chi_2^P$, we account only those components of corresponding tensors that make the main contribution to the dispersion and does not violate the circular shape of dispersion contours.

\section{Extraction of multipolar susceptibilities}

We have discussed that the knowledge of multipolar susceptibility tensors (see Eq.~\ref{eq:2}) allows us to formulate both the dispersion relation and boundary conditions. Nevertheless, we still need to extract these susceptibilities somehow. A number of methods are known in the literature to do this. One of the most popular and simple methods are the phenomenological approaches in the spirit of the Nicolson-Ross-Weir method~\cite{nicolson1970measurement,weir1974automatic,menzel2008retrieving}, which are based on the fitting of the reflection and transmission spectra. Such methods are extremely simple in application, but they can be mistakenly applied to inappropriate structures that fundamentally cannot be described in the scope of the considered model. This type of activity often lead to nonphysical and misleading results (see~\cite{simovski2018composite} for details).

Another approach is based on the multipolar decomposition of optical response of each metaatom and consideration of their coupling in the infinite lattice~\cite{simovski2018composite,petschulat2008multipole,gorlach2014effect,belov2005}. Such technique automatically frees us from the necessity to distinguish the contribution of different multipoles, but it might be technically complicated.

In this work, we apply the current-injection-based approach~\cite{silveirinha2007metamaterial,alu2011first}. As it has been already mentioned above, the main idea of such method is to inject the external current $\mathbf{j}^\mathrm{ext}(\omega,\mathbf{k}) \propto e^{i\mathbf{kr}}e^{-i\omega t}$ inside the infinite photonic crystal in order to excite it at arbitrary frequencies $\omega$ and $\mathbf{k}$. In turn, the obtained microscopic fields should be averaged to deal with macroscopic values. The original approach~\cite{silveirinha2007metamaterial} considers only the the generalized polarizability, which describes trustfully only dispersion of the bulk modes.

In some cases, knowledge of the generalized polarizability, $\mathbf{P}^\mathrm{gen}(\omega,\mathbf{k})$, is still enough for extracting multipolar susceptibilities.
For instance, sometimes we can fully attribute the spatial dispersion of generalized permittivity $\hat{\varepsilon}^\mathrm{gen}$ with dipole magnetization since it dominates at low frequencies~\cite{landau2013electrodynamics}. For instance, quadratic contributions to spatial dispersion is commonly attributed to the artificial magnetism and in turn, is described by effective permeability $\mu_\mathrm{eff}$. In this study we refer such approach to as $\varepsilon_\mathrm{eff}-\mu_\mathrm{eff}$ one. Nevertheless, this approach is fundamentally limited and can not be applied if we aim to go further and consider structures with a more complicated optical response.

For this purpose, here we propose a numerical approach to distinguish contributions of different multipolar susceptibilities. First, we apply dynamic averaging procedure not only for the electric dipole polarization, but for the high multipole moments as well:

\begin{align}
    \tilde{\mathbf{P}}^\mathrm{gen}(\omega,\mathbf{k})&=\frac{1}{-i\omega}\frac{1}{V}\int_V\mathbf{j}^\mathrm{micro}(\mathbf{r})e^{-i\mathbf{kr}}d^3\mathbf{r},\\
    \tilde{\mathbf{M}}^\mathrm{gen}(\omega,\mathbf{k})&=\frac{1}{2c}\frac{1}{V}\int_V\mathbf{r}\times\mathbf{j}^\mathrm{micro}(\mathbf{r})e^{-i\mathbf{kr}}d^3\mathbf{r},\label{eq:D2}\\
    \hat{\tilde{\mathbf{Q}}}^\mathrm{gen}(\omega,\mathbf{k})&=\frac{1}{-i\omega}\frac{1}{V}\int_V\left[\mathbf{r}\otimes\mathbf{j}^\mathrm{micro}+\mathbf{j}^\mathrm{micro}\otimes\mathbf{r}\right]e^{-i\mathbf{kr}}d^3\mathbf{r},\\
    \hat{\tilde{\mathbf{S}}}^\mathrm{gen}(\omega,\mathbf{k})&=\frac{2}{3c}\frac{1}{V}\int_V\left[\mathbf{r}\times\mathbf{j}^\mathrm{micro}\right]\otimes\mathbf{r}e^{-i\mathbf{kr}}d^3\mathbf{r}.
\end{align}

Obviously, when we average corresponding moments of microscopic polarization currents we obtain not the densities of corresponding multipole moments, but their "generalized" counterparts. The reason is that in the bulk medium it is still impossible to distinguish the magnetic dipole polarization from the appropriate gradient of magnetic quadrupole polarization. In other words, we can represent these generalized quantities as a series over the true densities of multipole moments:

\begin{align}
    \mathbf{P}^\mathrm{gen}(\omega,\mathbf{r})&=\mathbf{P}+\frac{i}{k_0}\nabla\times\mathbf{M}-\frac{1}{2}\nabla\cdot\hat{\mathbf{Q}}-\frac{i}{2 k_0}\nabla\times\hat{\mathbf{S}}\cdot\nabla+\dots,\label{eq:D5}\\
    \mathbf{M}^\mathrm{gen}(\omega,\mathbf{r})&=\mathbf{M}-\frac{1}{2}\hat{\mathbf{S}}\cdot\nabla+\dots,\label{eq:D6}\\
    \mathbf{Q}^\mathrm{gen}(\omega,\mathbf{r})&=\mathbf{Q}+\dots,\label{eq:D7}\\
    \mathbf{S}^\mathrm{gen}(\omega,\mathbf{r})&=\mathbf{S}+\dots,\label{eq:D8}
\end{align}
where Eq.~\ref{eq:D5} is the well known expansion~\cite{raab2005multipole,simovski2018composite,evlyukhin2016optical,silveirinha2007metamaterial,jackson1998classical} already mentioned above and Eq.~\ref{eq:D6}-\ref{eq:D8} are the similar expressions for densities of higher multipole moments. In Appendix~\ref{sec:magnetic_density} we derive the value of second term from Eq.~\ref{eq:D6}.

Nevertheless, we are ultimately interested in non-generalized quantities, so now we are able to find them from this system. Quadrupole moments can be approximated by their generalized values, since octupole moments are assumed to be small enough:

\begin{align}
    \mathbf{Q}&\approx\mathbf{Q}^\mathrm{gen},\\
    \mathbf{S}&\approx\mathbf{S}^\mathrm{gen}.
\end{align}

As for densities of dipole moments, in order to find them, we substitute the known expressions in Eq.~\ref{eq:D5}-\ref{eq:D6} and obtain the following connections:

\begin{align}
    \mathbf{P}^\mathrm{gen}&=\mathbf{P}+\frac{i}{k_0}\nabla\times\mathbf{M}^\mathrm{gen}-\frac{1}{2}\nabla\cdot\hat{\mathbf{Q}}^\mathrm{gen},\label{eq:D11}\\
    \mathbf{M}^\mathrm{gen}& = \mathbf{M}-\frac{1}{2}\hat{\mathbf{S}}^\mathrm{gen}\cdot\nabla.\label{eq:D12}
\end{align}

Finally, we easily express the true values of electric and magnetic dipole moment densities:

\begin{align}
    \mathbf{P}&=\mathbf{P}^\mathrm{gen}-\frac{i}{k_0}\nabla\times\mathbf{M}^\mathrm{gen}+\frac{1}{2}\nabla\cdot\hat{\mathbf{Q}}^\mathrm{gen},\label{eq:D13}\\
    \mathbf{M}& = \mathbf{M}^\mathrm{gen}+\frac{1}{2}\hat{\mathbf{S}}^\mathrm{gen}\cdot\nabla.\label{eq:D14}
\end{align}

In Appendix~\ref{sec:magnetic_density} we argue that these expressions are not approximate, but precise.

\section{Generalized density of magnetic dipole moment}
\label{sec:magnetic_density}

It has been already discussed that when we try to find averaged density of some mulipole moment, we inevitably account for the contribution of gradients of higher multipole moments, which are fundamentally indistinguishable in the bulk. In particular, the form of different contributions for the electric dipole polarization (Eq.~\ref{eq:D5}) is well known in the literature. In order to separate these contributions, we need to consider densities of other multipole moments and their representation in the form of a series as well. In particular, we are mostly interested in the "correction" to the density of magnetic dipole moment.

We calculate density of generalized magnetic dipole moment from the averaging of microscopic currents in accordance with Eq.~\ref{eq:D2}. In principle, we may try to decompose such an integral into the series of different contributions in a way demonstrated in~\cite{jackson1998classical} for the density of dipole moment. Nevertheless, it is much easier to employ the already developed expression for the generalized polarization (Eq.~\ref{eq:D5}).

Indeed, the generalized polarizability, $\mathbf{P}^\mathrm{gen}(\mathbf{r})$, unlike the ordinary one, $\mathbf{P}(\mathbf{r})$, describes all the currents that flow in the bulk. Moreover, a macroscopically averaged current 
\begin{equation}
    \mathbf{j}^\mathrm{macro}=-i\omega \mathbf{P}^\mathrm{gen} \label{eq:D15}
\end{equation}
should be non-distinguishable from the the microscopic currents when we consider them on a macroscopic scale.

Now, let us substitute the expression for the macroscopic current (Eq.~\ref{eq:D15}) into Eq.~\ref{eq:A2} and calculate the generalized magnetic dipole moment, $\mathbf{m}^\mathrm{gen}$, of some small volume $V$.
Since the magnetization is ultimately a macroscopic characteristic there should be no difference of whether we integrate over microscopic or macroscopic currents (their integral moments should not differ):

\begin{multline}
    \mathbf{m}^\mathrm{gen}= \frac{1}{2c}\int_V\mathbf{r}\times\mathbf{j}^\mathrm{micro}d^3\mathbf{r}= \frac{1}{2c}\int_V\mathbf{r}\times\mathbf{j}^\mathrm{macro}d^3\mathbf{r}=\\\frac{1}{2c}\int_V\mathbf{r}\times\left(-i\omega \mathbf{P}^\mathrm{gen}\right)d^3\mathbf{r}
\end{multline}

In turn, we can substitute four first terms of the series for generalized polarization from Eq.~\ref{eq:D5} and obtain the following:

\begin{multline}
    \mathbf{m}^\mathrm{gen}=  \frac{-ik_0}{2}\int_V\mathbf{r}\times\left( \mathbf{P}+\frac{1}{-ik_0}\nabla\times\mathbf{M}-\right.\\\left.\frac{1}{2}\nabla\cdot\hat{\mathbf{Q}}-\frac{1}{-2ik_0}\nabla\times\hat{\mathbf{S}}\cdot\nabla\right)d^3\mathbf{r}= \\\mathbf{m}^{\mathrm{gen},P}+\mathbf{m}^{\mathrm{gen},M}+\mathbf{m}^{\mathrm{gen},Q}+\mathbf{m}^{\mathrm{gen},S}.
\end{multline}

Now, we consider each of the four first terms separately. Term corresponding to electric polarization can be neglected since the integrand is proportional to the radius-vector $\mathbf{r}$, which is small at the zero point: 
\begin{equation}
     \mathbf{m}^{\mathrm{gen},P} = \frac{-ik_0}{2}\int_V\mathbf{r}\times\mathbf{P}d^3\mathbf{r}\approx 0
\end{equation}
For the second term we apply integration by parts:

\begin{multline}
     \mathbf{m}^{\mathrm{gen},M} = \frac{-ik_0}{2}\int_V\mathbf{r}\times\frac{1}{-ik_0}\nabla\times\mathbf{M}d^3\mathbf{r}=\\\frac{1}{2}\int_V\mathbf{r}\times\nabla\times\mathbf{M}d^3\mathbf{r}
\end{multline}

\begin{multline}
     m^{\mathrm{gen},M}_i = \frac{1}{2}\int_V \epsilon_{ijk} r_j\epsilon_{kmn}\partial_m M_n d^3\mathbf{r}=\\
     \frac{1}{2} \int_V \epsilon_{ijk} \epsilon_{kmn} r_j\partial_m M_n d^3\mathbf{r}=\frac{1}{2} \int_S \epsilon_{ijk} \epsilon_{kmn} r_j M_n n_mdS-\\\frac{1}{2} \int_V \epsilon_{ijk} \epsilon_{kmn} \delta_{jm} M_nd^3\mathbf{r} =\\ \int_S\dots dS - \frac{1}{2} \int_V \epsilon_{ijk} \epsilon_{kjn}  M_n d^3\mathbf{r} =\\
     \int_S\dots dS + \frac{1}{2} \int_V \epsilon_{kji} \epsilon_{kjn}  M_n d^3\mathbf{r} =\\ \int_S\dots dS + \frac{1}{2} \int_V 2\delta_{in}  M_n d^3\mathbf{r} =\\ \int_S\dots dS + \int_V  M_i d^3\mathbf{r}, \label{eq:E5}
\end{multline}
where $\mathbf{n}$ is a unit normal vector to the surface of the integration volume. 
As we can see, there is some additional surface contribution associated with surface multipole currents. Nevertheless, they does matter only when there is a physical interface of the considered metamaterial. Moreover, even in this case, the contribution of these surface currents can be accounted just by using appropriate boundary conditions, which constitutes the approach which we follow. Since our aim is to connect only the volume densities of multipole moments we do not care about the surface contributions.
As we can see, Eq.~\ref{eq:E5} confirms the obvious fact that the leading term for $\mathbf{M}^\mathrm{gen}$ is the ordinary magnetization, $\mathbf{M}$.

Now, let us consider the term associated with electric quadrupole contribution in the same manner:

\begin{multline}
         \mathbf{m}^{\mathrm{gen},Q}=  \frac{-ik_0}{2}\int_V\mathbf{r}\times\left( -\frac{1}{2}\nabla\cdot\hat{\mathbf{Q}}\right)d^3\mathbf{r} = \\
         \frac{ik_0}{4}\int_V\mathbf{r}\times\nabla\cdot\hat{\mathbf{Q}}d^3\mathbf{r}  
\end{multline}

\begin{multline}
         m^{\mathrm{gen},Q}_i = \frac{ik_0}{4}\int_V\epsilon_{ijk} r _j\partial_nQ_{nk}d^3\mathbf{r} =\\ = \frac{ik_0}{4}\int_S\epsilon_{ijk} r_jQ_{nk}n_n dS-\frac{ik_0}{4}\int_V\epsilon_{ijk} \delta_{jn}Q_{nk}d^3\mathbf{r}= \\  \int_S \dots dS-\frac{ik_0}{4}\int_V\epsilon_{ijk} Q_{jk}d^3\mathbf{r}=\\ \int_S \dots dS-\frac{ik_0}{4}\int_V\frac{1}{2}\left(\epsilon_{ijk} Q_{jk}+\epsilon_{ikj} Q_{kj}\right)d^3\mathbf{r}=\\ \int_S \dots dS-\frac{ik_0}{4}\int_V\frac{1}{2}\epsilon_{ijk}\left( Q_{jk}- Q_{kj}\right)d^3\mathbf{r}=\\ \int_S \dots dS - 0.
\end{multline}
As we can see, we proved the obvious fact that due to its symmetry, the electric quadrupole tensor does not make a contribution to the generalized magnetic dipole moment.

Finally, we consider the contribution of magnetic quadrupole moment:
    
\begin{multline}
         \mathbf{m}^{\mathrm{gen},S}=  \frac{-ik_0}{2}\int_V\mathbf{r}\times\left(-\frac{1}{-2ik_0}\nabla\times\hat{\mathbf{S}}\cdot\nabla\right)d^3\mathbf{r} =\\   -\frac{1}{4}\int_V\mathbf{r}\times\nabla\times\hat{\mathbf{S}}\cdot\nabla d^3\mathbf{r},   
\end{multline}
\begin{multline}
         m^{\mathrm{gen},S}_i= -\frac{1}{4}\int_V \epsilon_{ijk} r_j\epsilon_{kmn}\partial_m\partial_p S_{np} d^3\mathbf{r} = \\  -\frac{1}{4}\int_S \epsilon_{ijk} r_j\epsilon_{kmn} n_m\partial_p S_{np} dS +\frac{1}{4}\int_V \epsilon_{ijk} \delta_{jm}\epsilon_{kmn}\partial_p S_{np} d^3\mathbf{r} =\\  \int_S \dots dS -\frac{1}{4}\int_V \epsilon_{kji} \epsilon_{kjn}\partial_p S_{np} d^3\mathbf{r} =  \\  \int_S \dots dS -\frac{1}{4}\int_V 2\delta_{in} \partial_p S_{np} d^3\mathbf{r} =\\  \int_S \dots dS -\frac{1}{2}\int_V  \partial_p S_{ip} d^3\mathbf{r}  = \int_S \dots dS -\frac{1}{2}\int_V   \hat{\mathbf{S}}\cdot \nabla d^3\mathbf{r}.\label{eq:D10}
\end{multline}
From Eq.~\ref{eq:D10}, we can see that the gradient of magnetic quadrupole moment makes some non-zero contribution.

Gathering all together, there are surface and volume contributions $\mathbf{m}^{\mathrm{gen}}=\int_S\dots dS + \int_V \mathbf{M}^\mathrm{gen} d^3\mathbf{r}$ and the density can be found as follows:

\begin{equation}
    \mathbf{M}^\mathrm{gen} = \mathbf{M}-\frac{1}{2} \hat{\mathbf{S}}\cdot \nabla.
\end{equation}

Finally, it is worth to discuss the similarities and differences between the considered series and classical Taylor series. The main similarity is that the series for the electric polarization can be obtained as a Taylor decomposition of the Green function or delta function in the integrand. Therefore, in a latent form, numerical coefficients of the corresponding terms contain factors of $\frac{1}{n!}$.

The main difference is another approach for handling these series. For instance, when we deal with some function $f(x) = \sum_{n=0}^{\infty} \frac{x^n}{n!}f^{(n)}(0)$ in the form of Taylor expansion we can differentiate each term separately to find the expansion for the derivative of function $f'(x)= \sum_{n=1}^{\infty}= \sum_{n=0}^{\infty}\frac{nx^{n-1}}{n!}f^{(n)}(0)= \sum_{n=1}^{\infty}\frac{x^{n-1}}{(n-1)!}f^{(n)}(0)= \sum_{n=0}^{\infty}\frac{x^{n}}{n!}f^{(n+1)}(0)= \sum_{n=0}^{\infty}\frac{x^{n}}{n!}(f')^{(n)}(0)$. It is important that during the differentiation of each term different values of the coefficient $n$ are dropped out, which results in a rise of a brand new series, which naturally can not be expressed in terms of the former one, which is in accordance with the fact that derivative of function at some point is generally independent on the original function at this point.

The different situation is observed, when we try to conduct a similar procedure for the generalized polarization series. In this case, in order to obtain the expression for the density of multipole moment of the order $n$, we need to multiply the series by the radius vector in corresponding power and integrate the expression $n$ times by parts. Additional numerical coefficients can drop out only when we take derivative of the radius vector during integration by parts. Moreover, all the terms are elaborated in the same way and therefore, the same numerical coefficient should drop out of them and the relationship between the terms in a new series would be the same as in the original one. This makes it possible to express one series in terms of the other one in a closed form. For instance, if we factor common curl out of the brackets in the Eq.~\ref{eq:D5} then we will discover the series for the generalized magnetic moment inside of them. The same idea should be valid for the electric quadrupole moment, which is also the multipole moment of the first order. Since there are no other multipole moments of this order, we can expect that the relation (Eq.~\ref{eq:D11}) is strict:

\begin{multline}
    \mathbf{P}^\mathrm{gen}=\mathbf{P}+\frac{1}{-ik_0}\nabla\times\mathbf{M}-\frac{1}{2}\nabla\cdot\hat{\mathbf{Q}}-\frac{1}{-2ik_0}\nabla\times\hat{\mathbf{S}}\cdot\nabla+\dots=\\
    \mathbf{P}+\frac{1}{-ik_0}\nabla\times\left(\mathbf{M}-\frac{1}{2}\hat{\mathbf{S}}\cdot\nabla+\dots\right)-\frac{1}{2}\nabla\cdot\left(\hat{\mathbf{Q}}+\dots\right)=\\
    \mathbf{P}+\frac{1}{-ik_0}\nabla\times\mathbf{M}^\mathrm{gen}-\frac{1}{2}\nabla\cdot\hat{\mathbf{Q}}^\mathrm{gen}.
\end{multline}
The same is valid for Eq.~\ref{eq:D12}.

\section{Calculation of true multipolar densities}

Now, everything is ready to calculate multipolar susceptibilities for our practical case. As it was already discussed, in order to excite electro-magnetic fields, we inject external current $\mathbf{j}^\mathrm{ext}\propto e^{i\mathbf{kr}}$. In particular, here we demonstrate the results for $k_x$ harmonic of $y$-directed current:

\begin{equation}
    \mathbf{j}^\mathrm{ext} = \frac{-i\omega}{4\pi}\begin{pmatrix}
        0\\1\\0
    \end{pmatrix}e^{ik_x x}\cdot 1(G).
\end{equation}
Although constant factor $\frac{-i\omega}{4\pi}\cdot1(G)$ might be arbitrary, we have specifically chosen it similar to the polarization current introduced by plane wave with $E_0=B_0=1(G)$ in some scatterer. Such choice makes most of the considered fields real-valued and their values are comparable with unity in Gaussian system.

\begin{figure}
    \centering
    \includegraphics[width=\linewidth]{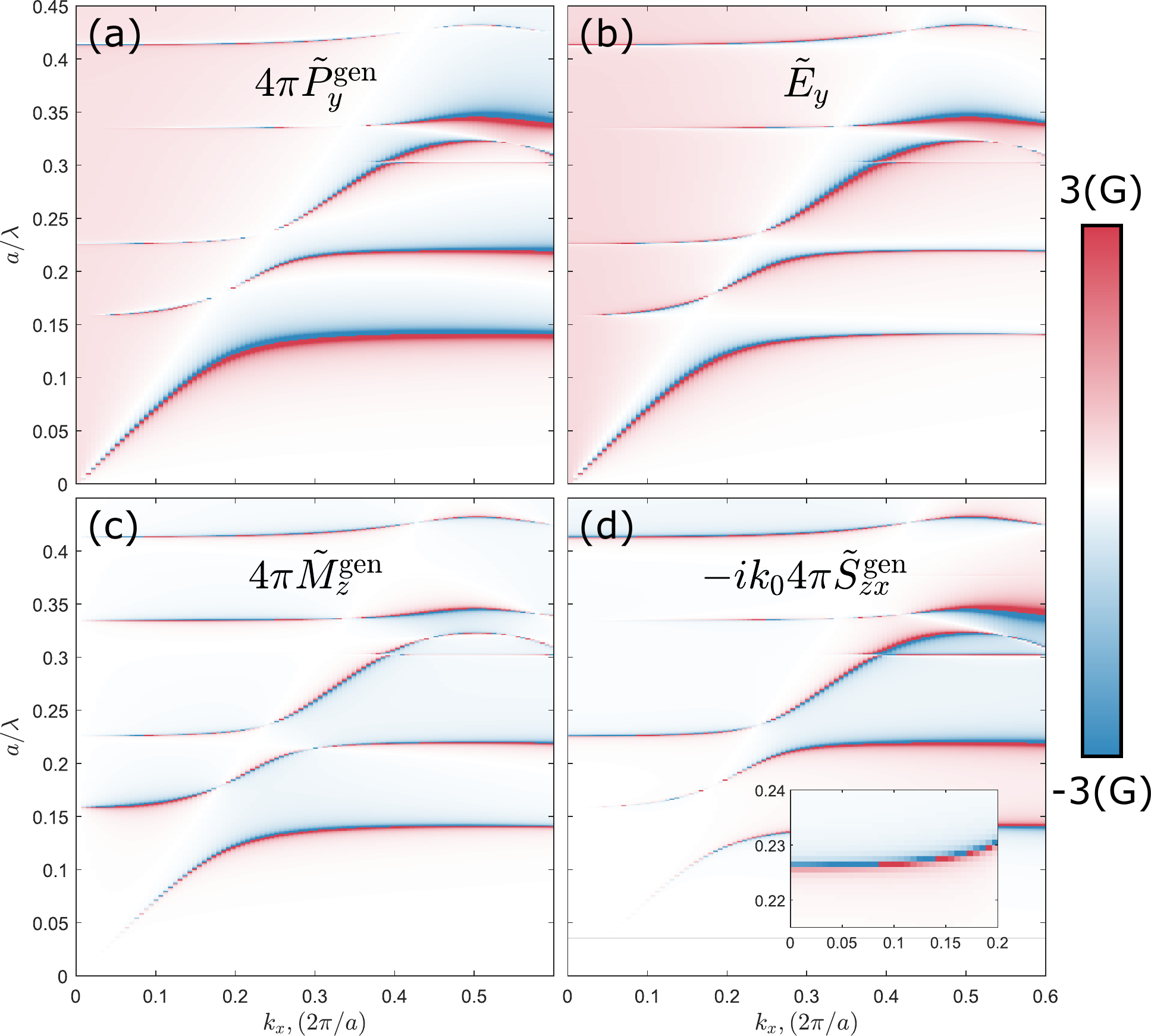}
    \caption{Spectral maps of generalized multipole densities, $\tilde{P}^\mathrm{gen}$ (panel (a)), $\tilde{M}^\mathrm{gen}$ (panel (c)), $\tilde{S}^\mathrm{gen}$ (panel (d)), and macroscopic electric field, $\tilde{E}$ (panel (b)) excitation by external current $j_y\propto e^{ik_xx-i\omega t}$.}
    \label{fig:fig5}
\end{figure}

Solution of the emission problem in COMSOL Multyphysics provides us with amplitudes of macroscopic electric field and densities of generalized multipole moments shown in Fig.~\ref{fig:fig5}. All of them demonstrate divergence of the fields in the vicinity of the bulk eigenmodes of metamaterial. In particular, in the low-energy limit, we observe a linear dispersion corresponding to the classical light cone of dielectric material. In turn, at higher energies there are multiple points of avoided crossing between the light cone and dispersionless Mie resonances of cylinders. Already at this stage we see that the magnetic quadrupole moment can be excited by electric field (without any gradient). Indeed, the inset of panel~(d) shows that at the $\Gamma$-point magnetic quadrupole moment experinces resonance, whereas all the gradients of macroscopic electric field are zero.

\begin{figure*}
    \centering
    \includegraphics[width=0.8\linewidth]{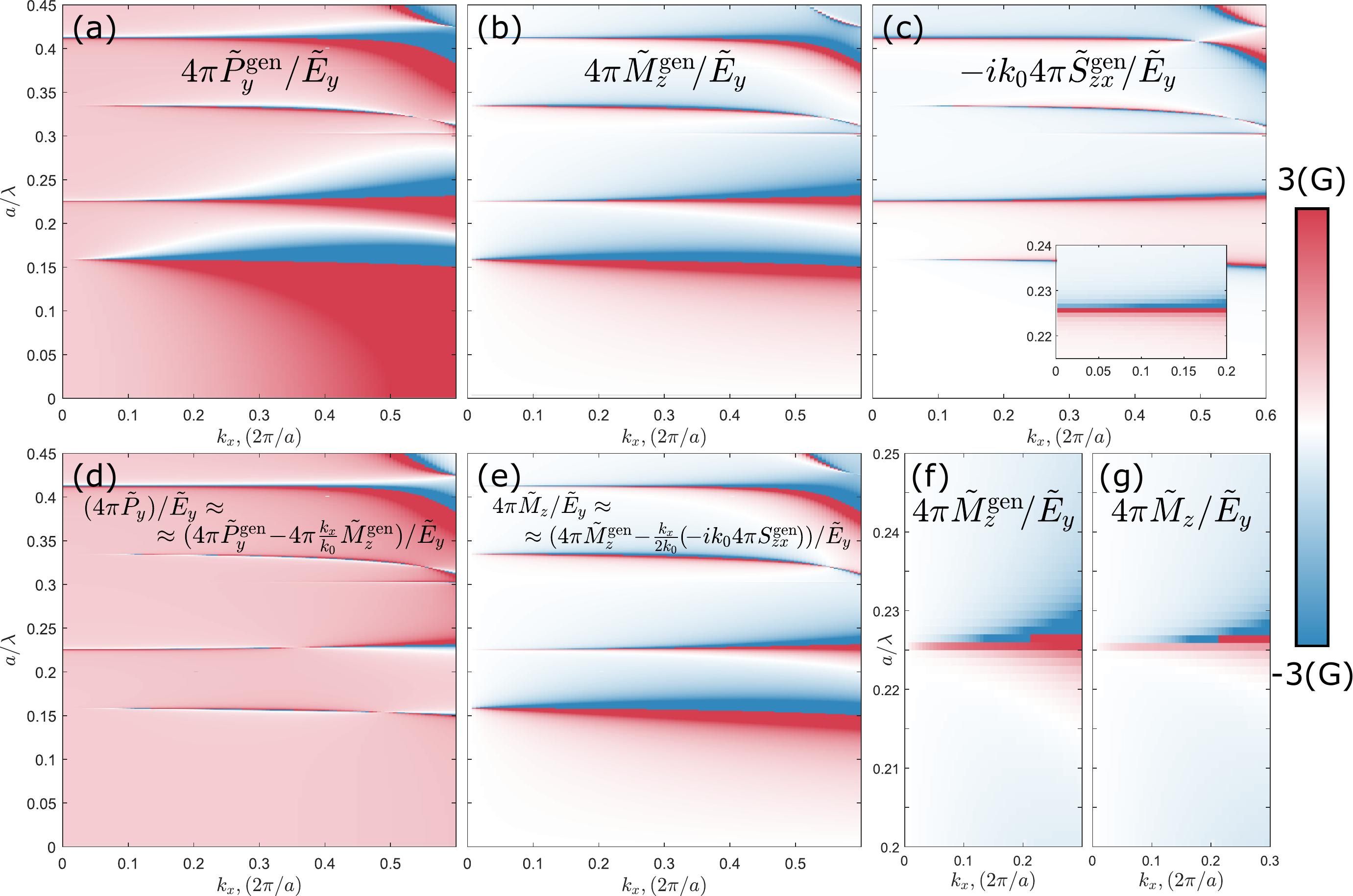}
    \caption{ (a-c) Spectral maps showing ratio of generalized multipole densities, $\tilde{P}^\mathrm{gen}$, $\tilde{M}^\mathrm{gen}$, $\tilde{S}^\mathrm{gen}$, and macroscopic electric field, $\tilde{E}$ demonstrate optical response of corresponding multipoles. (d-e) Spectral maps showing ratio of multipole densities, $\tilde{P}$, $\tilde{M}$, and macroscopic electric field, $\tilde{E}$. Spatial dispersion for electric dipole polarization in panel~(d) is strongly suppressed in comparison with generalized one in panel~(a). Panels (f-g) show that the optical response of generalized and ordinary magnetization near the magnetic quadrupole Mie resonance are also different, which is associated with magnetic quadrupole moment contribution. }
    \label{fig:fig6}
\end{figure*}

Nevertheless, we are ultimately interested in the optical response of metamaterial, so we consider a ratio of generalized mulipole densities and macroscopic electric field in Fig.~\ref{fig:fig6}~(a-c). Since both multipole densities and electric field experience resonance at the same points, their ratio is expected to be at least finite. Indeed, as we can see from Fig.~\ref{fig:fig6}~(a-c), the electric dipole, magnetic dipole and magnetic quadrupole responses of metamaterial are smooth and continuous at the points corresponding to the dispersion of eigenmodes. Concurrently, they demonstrate almost dispersionless resonances at a number of frequencies corresponding to the Mie modes of metaatoms. From the inset of panel~(c) we again clearly observe excitation of magnetic quadrupole moment by electric field, but not its gradients, ($\mathbf{k}=0$) at the second Mie resonance.

At the same time, we should note that, except of the resonance-determined frequency dependence of optical responses, there is also smooth dependence on the $\mathbf{k}$-vector, which is known as spatial dispersion. This dependence, might be associated either with excitation of higher order multipole moments or with the optical response on the gradients of the electric field.
In order to extract multipolar susceptibilities from Eq.~\ref{eq:sus1}-\ref{eq:sus3}, we should distinguish these contributions and proceed from generalized densities of multipole moments to the ordinary ones. For this purpose, we apply Eq.~\ref{eq:D13}-\ref{eq:D14} for our particular case of a plane wave and obtain:

\begin{align}
    \tilde{P}_y &=    \tilde{P}^\mathrm{gen}_y - \frac{k_x}{k_0}\tilde{M}^\mathrm{gen}_z,\\
    \tilde{M}_z &=    \tilde{M}^\mathrm{gen}_z + \frac{i k_x}{2}\tilde{S}^\mathrm{gen}_{zx}.    
\end{align}

The optical response of corresponding quantities $\tilde{P}_y$ and $\tilde{M}_z$ is shown in Fig.~\ref{fig:fig6}~(d-e). From the comparison of panels (a) and (d) we clearly see that there is almost no spatial dispersion for $\tilde{P}_y$ in contrast with $\tilde{P}^\mathrm{gen}_y$, which is a proof that most of the corresponding effects are associated with the contribution of magnetic dipole moment (artificial magnetism) and magnetic quadrupole one. At the same time, the remaining $\mathbf{k}$-dependence associated with $\chi_2^P$ as well as the main factor, $\chi_0^P$,  can be easily extracted from the Tailor expansion of $\tilde{P}_y/\tilde{E}_y$ function at the $\Gamma$ point ($\mathbf{k}=0$).

Speaking about the magnetization, the difference between $\tilde{M}_Z^\mathrm{gen}$ and $\tilde{M}_z$, is not so obvious from the comparison of panels~(b)~and~(d). Nevertheless, it can be noted by accurately considering area of the second, magnetic quadrupole Mie-resonance in Fig.~\ref{fig:fig6}~(f-g). As we have already discussed in the article, although the contribution of magnetic quadrupole moment resonance is significant only in a narrow frequency band, in this band it substantially affects the overall behavior. Finally, susceptibilities, $\chi_1^M$, and, $\chi_0^s$, are extracted from the Tailor expansion of functions $\tilde{M}_z/\tilde{E}_y$ and $\tilde{S}_{zx}/\tilde{E}_y$ from panels (e) and (c).

\section{Boundary conditions}

\label{sec:boundary_conditions}

Knowledge of multipolar susceptibilities allows us not only to determine dispersion of the bulk waves, but also formulate appropriate boundary conditions. In this section, we reduce boundary conditions from~\cite{graham2000multipole} for our particular case. We transform the formulas from SI (International System of Units) to Gaussian units, employ our notations, consider $x=\mathrm{const}$ interface, and drop out all the multipole moments and their components that are out of our consideration. Accounting for the magnetic quadrupole, electric and magnetic dipoles allows us to obtain conditions on tangential fields in the following form: 
\begin{multline}
B_z-4\pi M_z+\\2\pi\left(\partial_x S_{zx}+\partial_y S_{zy}+\partial_z S_{zz}-\partial_z S_{xx}\right)=\mathrm{const},
\end{multline}
\begin{equation}
E_y+2\pi ik_0 S_{zx}=\mathrm{const}.
\end{equation}
For our particular model $S_{xx}=S_{zz}=0$, boundary conditions can be slightly simplified: 

\begin{align}
B_z-4\pi M_z+2\pi\left(\partial_x S_{zx}+\partial_y S_{zy}\right)=\mathrm{const},\\
E_y+2\pi ik_0 S_{zx}=\mathrm{const}.
\end{align}

In turn, we are able to substitute constitutive relations from Eq.~\ref{eq:sus1}-\ref{eq:sus3}:

\begin{align}
    B_z(1-4\pi\chi_1^M)-\frac{2\pi i}{k_0}\chi_0^S(-\partial_xE_y+\partial_yE_x)=\mathrm{const},\\
    \left(1-2\pi\chi_0^S \right) E_y=\mathrm{const}.
\end{align}
And finally express the boundary conditions in a convenient form:

\begin{align}
    (1-4\pi\chi_1^M-2\pi\chi_0^S) B_z=\mathrm{const},\\
    \left(1-2\pi\chi_0^S \right) E_y=\mathrm{const}.
\end{align}
For the sake of convenience, we introduce effective parameters:

\begin{align}
 \frac{B_z}{\mu}=\mathrm{const},\\
    \frac{E_y}{\eta}=\mathrm{const}.
\end{align}

\section{Generalization of Fresnel Equations}
\label{sec:fresnel}

Now, having derived the boundary conditions, we are able to employ them and update the Fresnel equations. In our case, we consider only $p$ polarization, which is described by the introduced model. We assume that the light wave comes from the 1${}^\mathrm{st}$ medium at $\Theta_1$ angle to normal and is refracted in the $2^\mathrm{nd}$ one at $\Theta_2$ angle. In such case we deal with following boundary conditions:

\begin{align}
     \frac{B^i+B^r}{\mu_1}&=\frac{B^t}{\mu_2},\\
     \frac{E^i \cos \Theta_1-E^r\cos \Theta_1}{\eta_1}&=\frac{E^t\cos \Theta_2}{\eta_2},
\end{align}
where $E^{i,r,t}$, and $B^{i,r,t}$ are amplitudes of electric and magnetic fields of incident, reflected and transmitted waves correspondingly. 
Taking into account that $B=nE$, we obtain the system of equations on reflection, $r$, and transmission, $t$, coefficients:

\begin{align}
     \frac{n_1}{\mu_1}(1+r)&=\frac{n_2}{\mu_2}t,\\
     \frac{ \cos \Theta_1}{\eta_1}(1-r)&=\frac{\cos \Theta_2}{\eta_2}t,
\end{align}
and finally solve it:

\begin{align}
     t &= \frac{2\mu_1n_2\eta_2 \cos \Theta_1}{\mu_1n_2\eta_2\cos \Theta_1+\mu_2n_1\eta_1\cos \Theta_2}\frac{\mu_2 n_1}{\mu_1 n_2},\\
    r &= \frac{\mu_1n_2\eta_2 \cos \Theta_1-\mu_2n_1\eta_1\cos \Theta_2}{\mu_1n_2\eta_2\cos \Theta_1+\mu_2n_1\eta_1\cos \Theta_2}.
\end{align}

These equations can be already used in practice if we remember that the propagation angles are connected via Snell's law $n_1\sin \Theta_1 =n_2\sin \Theta_2$. Concurrently, we can represent the equations in another common form via normal components of wavevector $k_{x,1/2}=\sqrt{n_{1/2}^2k_0^2-k_y^2} = n_{1/2}k_0\cos \Theta_{1/2}$:

\begin{align}
     t &= \frac{2\mu_1n^2_2\eta_2 k_{x1}}{\mu_1n^2_2\eta_2 k_{x1}+\mu_2n^2_1\eta_1 k_{x2}}\frac{\mu_2 n_1}{\mu_1 n_2},\\
    r &= \frac{\mu_1n^2_2\eta_2 k_{x1}-\mu_2n^2_1\eta_1 k_{x2}}{\mu_1n^2_2\eta_2 k_{x1}+\mu_2n^2_1\eta_1 k_{x2}}.
\end{align}

\section{ Reflection from metamaterial interface}
\label{sec:reflection}

\begin{figure}
    \centering
    \includegraphics[width=\linewidth]{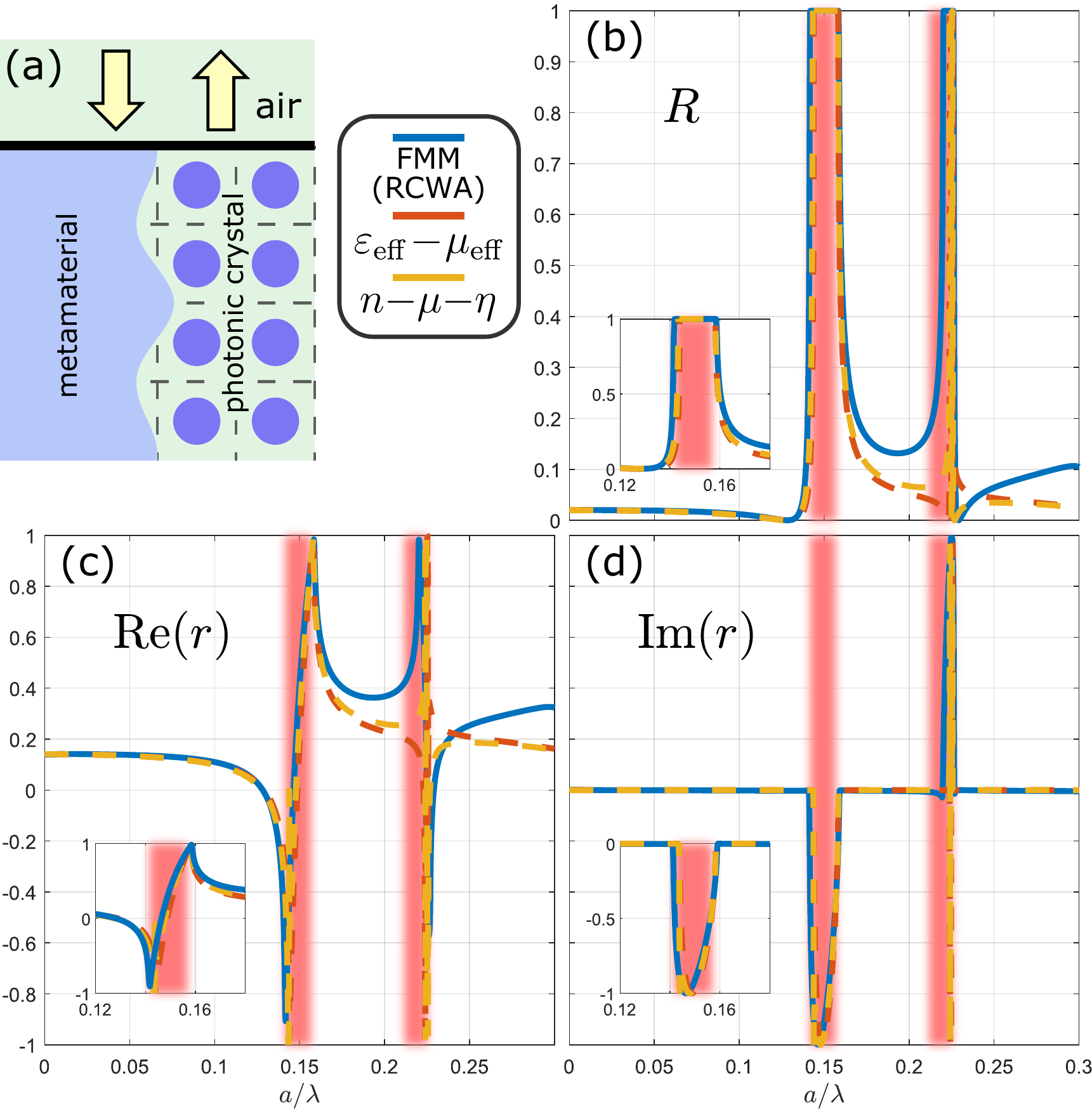}
    \caption{ (a) Schematic of normally incident light reflection from semi-infinite metamaterial. Panels (b,c,d) demonstrate the reflection spectrum and spectra of real and imaginary parts of the amplitude reflection coefficient. Blue lines correspond to reference calculations conducted for photonic crystal structure via the Fourier modal method~\cite{tikhodeev2002}. Dashed red lines are calculated in $\varepsilon_\mathbf{eff}-\mu_\mathrm{eff}$ metamaterial approximation, which fully attributes quadratic  spatial dispersion to effective permeability. Dashed yellow lines correspond to developed in this study $n-\mu-\eta$ metamaterial approximation, which accounts for magnetic qudrupole as well. Both approximations well describe the first, magnetic dipole Mie resonance. Concurrently, yellow curve in contrast with the red one demonstrates qualitatively a correct behavior near the second, magnetic quadrupole Mie resonance (see panel~(c)).  }
    \label{fig:fig7}
\end{figure}

Now, when we have derived Fresnel equations, we can apply them to calculate the reflection coefficient from the interface of metamaterial. In order to verify our results, we consider reflection from semi-infinite metamaterial as shown in Fig.~\ref{fig:fig7}~(a). The reference spectrum depicted by the blue lines in Fig~\ref{fig:fig7}~(b-d) is calculated via the Fourier modal method for the actual photonic crystal structure, while the red lines show the results obtained in $\varepsilon_\mathrm{eff}-\mu_\mathrm{eff}$ approximation, which accounts only for the dipole moments densities and the yellow lines show the spectra calculated via the proposed $n-\mu-\eta$ approach, which additionally accounts for the contribution of magnetic quadrupole moment. The light-red areas correspond to the frequency bands where the metamaterial approximation is expected to perform poorly. It is associated either to proximity of the dispersion to the boundary of the Brillouin zone or very fast decay of evanescent fields inside the photonic stop-band.

Nevertheless, both versions of effective medium approximations almost perfectly describe reflection not only for the lowest energies, but even in the first photonic stop-band associated with the magnetic dipole Mie resonance. What is especially important, they correctly describe not only the reflection intensity (panel (b)), but also the real and imaginary parts of the amplitude reflection coefficient (panels (c-d)). A correct amplitude and phase of reflection together with appropriate dispersion of bulk waves implies that effective medium approximation can be legally used to describe complicated structures comprised of corresponding metamaterial.

For the energies above the first resonance ($a/\lambda>0.17$), both approaches deviate from the true behavior. The main reason is a necessity to account for the spatial dispersion effects of the 4th order. However, even in the framework of the 2nd order approximation, $n-\mu-\eta$ performs better than $\varepsilon_\mathrm{eff}-\mu_\mathrm{eff}$. The main difference is qualitative behavior in the vicinity of the second resonance. It is the yellow line that follows the blue one in panel~(c), i.e. rises to infinity below the resonant energy and comes back from negative infinity above it. Even such a small feature allows to describe the behavior of metamaterial at the bottom of the opening of transmission band in much more consistent way.


%

\end{document}